\pdfoutput=1
\documentclass[DIV12]{scrartcl}
\usepackage[utf8]{inputenc}
\usepackage[T1]{fontenc}
\usepackage[english]{babel}
\usepackage{authblk}
\usepackage{cite}
\usepackage{graphicx}
\usepackage{amsmath}
\usepackage[squaren,Gray,textstyle,thinqspace]{SIunits}
\setkomafont{disposition}{\normalcolor\rmfamily}

\setcapindent{0mm}
\typearea[current]{current}
\begin{document}
\title{Airy beam induced optical routing}
\author{Patrick Rose, Falko Diebel, Martin Boguslawski, and Cornelia Denz}
\affil{\large Institut für Angewandte Physik and Center for Nonlinear Science (CeNoS), Westfälische Wilhelms-Universität Münster,\\ Corrensstraße~2/4, 48149~Münster, Germany}
\date{}
\maketitle
\begin{abstract}
We present an all-optical routing scheme based simultaneously on optically induced photonic structures and the Airy beam family. The presented work utilizes these accelerating beams for the demonstration of an all-optical router with individually addressable output channels. In addition, we are able to activate multiple channels at the same time providing us with an optically induced splitter with configurable outputs. The experimental results are corroborated by corresponding numerical simulations.
\end{abstract}
%
%
The vision to overcome the unavoidable bandwidth constraints of electronics by heading towards an all-optical computing architecture is based on the idea of light guiding light. Following this concept, we present a scheme for the adaptive spatial routing of light using optically induced photonic structures.

During the last decade, the optical induction technique~\cite{ref:efremidis2002} got a lot of attention. This approach allows for the preparation of a multitude of linear and nonlinear refractive index schemes by illuminating a photorefractive material with structured intensity distributions. Over the years, the achievable lattice complexity developed from fundamental one- and two-dimensional patterns~\cite{ref:fleischer2003_1,ref:fleischer2003_2} to three-dimensional quasi-crystals~\cite{ref:xavier2010} and even the multiplexing of arbitrary structures~\cite{ref:boguslawski2012} is possible nowadays.

In this respect, optically induced complex two-dimensional structures are particularly interesting since they require a modulation in the transverse dimensions but remain invariant in a third direction. Certainly, the inducing intensity distribution has to fulfill this requirement as well. While this at first seems to contradict the ubiquitous diffraction, Durnin et~al.\ demonstrated the realization of such a nondiffracting beam in 1987~\cite{ref:durnin1987} and his discovery today is called Bessel beam. Besides Bessel beams, other families of nondiffracting beams were found over the years and now it is well-known that four different families~-- Bessel, Mathieu, Weber, and discrete nondiffracting beams~-- exist~\cite{ref:gutierrez-vega2000,ref:bandres2004}. Moreover, we recently demonstrated that all these different families can even be used for the optical induction of corresponding photonic lattices~\cite{ref:rose2012}.

In addition to these classical nondiffracting beams, the related class of accelerating beams is a field of vivid research as well. During propagation, the intensity distribution of an accelerating beam retains its shape but it is shifted in the transverse plane leading altogether to an accelerated propagation trajectory~\cite{ref:bandres2009}. Based on solutions to the Schrödinger equation found by Berry and Balazs~\cite{ref:berry1979}, the first accelerating optical beam was described and realized in 2007~\cite{ref:siviloglou2007_1,ref:siviloglou2007_2}. This so-called Airy beam has already been utilized for many different applications ranging from abruptly autofocusing waves~\cite{ref:efremidis2010} to optical snowblowers in micromanipulation experiments~\cite{ref:baumgartl2008}. Moreover, recent works introduced further ideas for controlling the trajectory of accelerated beams~\cite{ref:liu2011,ref:efremidis2011,ref:chavez-cerda2011,ref:ye2011} and some of them support even arbitrary paths~\cite{ref:greenfield2011,ref:froehly2011}.

\begin{figure}[htb]
    \centering
    \includegraphics[width=9.5cm]{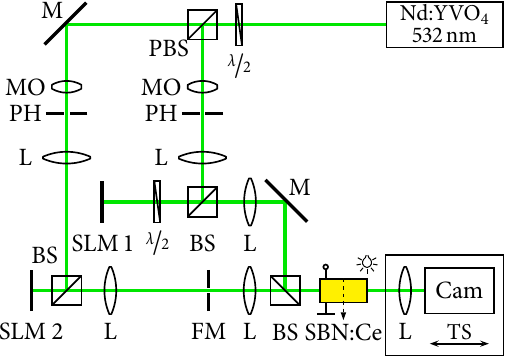}
    \caption{Schematic experimental setup for Airy beam induced optical routing. FM:~Fourier mask, L:~lens, M:~mirror, MO:~microscope objective, (P)BS:~(polarizing) beam splitter, PH:~pinhole, SLM:~spatial light modulator, TS:~translation stage.}
    \label{fig:setup}
\end{figure}
In the following, we introduce an adaptive all-optical routing technique based on these beams. The underlying experimental setup is schematically shown in Fig.~\ref{fig:setup}. The beam from a frequency-doubled Nd:YVO$_{4}$ laser at a wavelength of \unit{532}{\nano\meter} is divided into two separate beams and each of them illuminates a programmable spatial light modulator~(SLM). The first modulator (SLM~1, see Fig.~\ref{fig:setup}) is used to imprint a cubic phase onto the incident beam. In combination with a Fourier transforming lens, this leads to the generation of an ordinarily polarized two-dimensional Airy beam~\cite{ref:siviloglou2007_2} at the front face of the depicted \unit{20}{\milli\meter} long photorefractive Sr$_{0.60}$Ba$_{0.40}$Nb$_{2}$O$_{6}$~(SBN:Ce) crystal. The crystal is externally biased with a dc electric field directed along its optical axis in order to allow the optical induction of a corresponding refractive index distribution. In addition, we can illuminate the crystal homogeneously with a white light source to erase any written refractive index configuration. Using a technique similar to the one described in~\cite{ref:terhalle2009}, the second SLM (cf.\ Fig.~\ref{fig:setup}) generates and positions an extraordinarily polarized input beam for the optically induced structure. Finally, an imaging lens and a camera mounted on a translation stage can be used to analyze the intensity distribution in different transverse planes.

\begin{figure}[htb]
    \centering
    \includegraphics[width=12cm]{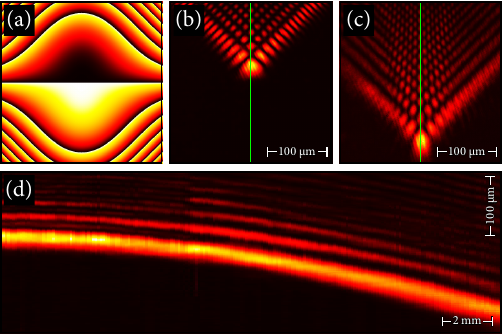}
    \caption{Two-dimensional Airy beam. (a)~Cubic spatial phase spectrum, (b)~experimentally observed Airy beam intensity distribution at the front face and (c)~at the back face of the crystal, (d)~accelerating intensity profile in longitudinal direction. The green lines in~(b) and~(c) mark the longitudinal plane shown in~(d). Besides~(a) all figures are normalized.}
    \label{fig:airy2d}
\end{figure}
The propagation characteristics of a two-dimensional Airy beam generated in the described setup are summarized in Fig.~\ref{fig:airy2d}. The cubic spatial phase spectrum put on the first SLM is shown in Fig.~\ref{fig:airy2d}a. After the Fourier transforming lens, we get the two-dimensional Airy beam depicted in Fig.~\ref{fig:airy2d}b at the front face of the crystal. The beam propagates through the photorefractive material, and at the crystal's back face we find the shifted two-dimensional Airy pattern shown in Fig.~\ref{fig:airy2d}c. While the outer Airy lobes already show some diffraction, the beam maximum is still extremely pronounced.

Moreover, we observe the transverse intensity pattern at  multiple longitudinal positions in between by shifting camera and optics with the translation stage. After stacking all these acquired two-dimensional images, we get a three-dimensional intensity distribution of the propagating Airy beam which allows for an analysis of the accelerated propagation dynamics in the longitudinal direction. Figure~\ref{fig:airy2d}d illustrates the undisturbed parabolic propagation of the Airy beam maximum in our experiment.

\begin{figure}[htb]
    \centering
    \includegraphics[width=12cm]{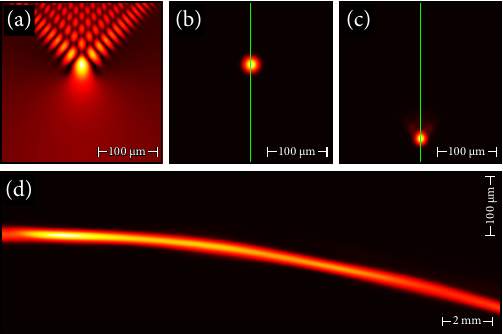}
    \caption{Numerical simulation of Airy beam induced guiding of light. (a)~Optically induced transverse refractive index modulation for the inducing beam configuration shown in Fig.~\ref{fig:airy2d}b, (b)~Gaussian input beam at the front face of the photorefractive crystal, (c)~output of the guided beam at the back face of the crystal, (d)~intensity distribution of the guided beam in longitudinal direction during propagation. The green lines in~(b) and~(c) mark the longitudinal plane shown in~(d). All figures are normalized.}
    \label{fig:simulation}
\end{figure}

If now the external dc field is applied to the photorefractive medium, the optical induction leads to a reversible refractive index modulation inside the material depending on the incident intensity distribution. The induced refractive index profile can be calculated using the so-called full anisotropic model~\cite{ref:zozulya1998}. In Fig.~\ref{fig:simulation}a, we show the numerically obtained refractive index modulation at the beginning of the SBN crystal for the inducing beam configuration shown in Fig.~\ref{fig:airy2d}. All numerical parameters are chosen to match the experimental conditions. Since the shape of the transverse intensity distribution of the Airy beam remains almost invariant during propagation (cf.\ Figs.~\ref{fig:airy2d}b and~\ref{fig:airy2d}c, respectively), the shape of the induced refractive index modulation is similar in every plane as well. Only the overall pattern is shifted following the acceleration of the inducing Airy beam. In this way, we get a curved path of increased refractive index going from the input position of the Airy beam maximum at the front face to the corresponding output position at the back.

This curved waveguide can then be used for the optical guiding of light. Using a split-step Fourier propagation method~\cite{ref:agrawal1995}, we simulate the linear propagation of a Gaussian beam (Fig.~\ref{fig:simulation}b) positioned at the maximum of the refractive index distribution (cf.\ Fig.~\ref{fig:simulation}a). The results are summarized in Figs.~\ref{fig:simulation}c and~\ref{fig:simulation}d. The Gaussian input is nicely guided to the addressed output position (Fig.~\ref{fig:simulation}c) and the intensity distribution of the beam in longitudinal direction (Fig.~\ref{fig:simulation}d) confirms as well that the beam follows exactly the accelerated path of the Airy beam used for the optical induction. These numerical results motivate the experimental implementation of an Airy beam induced optical routing setup.

\begin{figure}[htb]
    \centering
    \includegraphics[width=12cm]{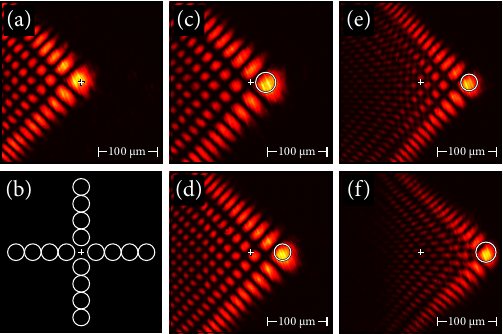}
    \caption{Output scheme of the Airy beam induced optical router. (a)~Input position of the Airy beam at the front face of the crystal, (b)~schematic array of 16 output channels given by four \unit{90}{\degree} beam rotations and four different accelerations each, (c),~(d),~(e),~(f)~output position of the Airy beam at the back face of the crystal for different accelerations. Besides~(b) all figures are normalized.}
    \label{fig:outputscheme}
\end{figure}

Since in our setup the Airy beam generation is based on a computer-controlled SLM, we can easily change the propagation characteristics. On the one hand, we can rotate the whole structure by simply rotating the phase pattern (cf.\ Fig~\ref{fig:airy2d}a) on the SLM. On the other hand, a stretching or compression of the cubic phase distribution leads to a change in the Airy beam's acceleration~\cite{ref:chremos2011} and therewith to different transverse shifts between the input position of the maximum at the front face and the output position at the back face of the crystal. Figure~\ref{fig:outputscheme} shows that in this way a combination of just four different orientations with four accelerations provides already 16 distinct output channels accessible from a single input position.

In Fig.~\ref{fig:outputscheme}a, the input position of an Airy beam turned by \unit{90}{\degree} with respect to the configuration shown in Fig.~\ref{fig:airy2d} is marked with a cross. Figures~\ref{fig:outputscheme}c--\ref{fig:outputscheme}f then highlight the variable shift between this input position and the output position of the Airy beam at the back face of the crystal provided by different beam accelerations. Combining these four curvatures with four \unit{90}{\degree} beam rotations gives us the configuration of possible outputs shown schematically in Fig.~\ref{fig:outputscheme}b.

Together with an external dc field, each of the two-dimensional Airy beams optically induces a different reversible refractive index structure that guides light from the central input position to the corresponding output channel. Furthermore, the induced structure can be erased again using the homogeneous white light illumination. Afterwards, the system is prepared for the optical induction of a new input-output configuration.

\begin{figure}[htb]
    \centering
    \includegraphics[width=12cm]{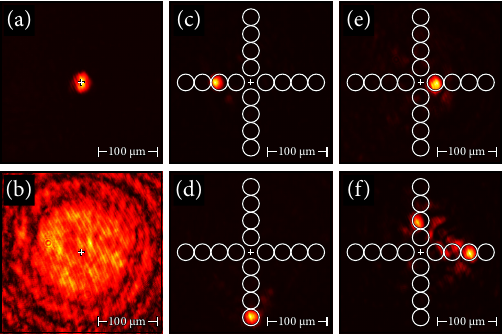}
    \caption{Airy beam induced optical routing. (a)~Gaussian input beam at the front face of the crystal, (b)~diffracted Gaussian output at the back face of the crystal without induced refractive index structure, (c),~(d),~(e)~guided output for different selected channels, (f)~optically induced splitter with multiple selected outputs. All figures are normalized.}
    \label{fig:routing}
\end{figure}

Now, we generate a low intensity Gaussian probe beam (cf.\ Fig.~\ref{fig:routing}a) as an input to our experimental system. Without an optically induced structure, the Gaussian beam diffracts and leads to the broadened output at the back face of the crystal shown in Fig.~\ref{fig:routing}b. However, if a single Airy beam induced refractive index channel is present, we observe a precisely guided beam at the addressed output position. Figures~\ref{fig:routing}c--\ref{fig:routing}e illustrate this for three arbitrarily chosen channels. The images were captured immediately after the launch of the Gaussian beam in order to observe definitely the linear propagation profile and the Airy beam is blocked during the recording.

Moreover, we can apply the idea of incremental multiplexing of optically induced structures~\cite{ref:rose2008} in order to activate multiple output channels at the same time. Utilizing the computer-controlled SLM, we can continuously switch between different Airy beam configurations faster than the time constants of the medium. The induced refractive index distribution then corresponds to the combination of the multiplexed beams. In this way, our approach is capable of inducing refractive index structures that simultaneously guide light from one input to multiple selected outputs. Figure~\ref{fig:routing}f demonstrates this unique feature for a two-channel splitter that divides the input signal (cf.\ Fig.~\ref{fig:routing}a) and guides it to the two arbitrarily chosen outputs.

In summary, we introduced an all-optical routing setup based on Airy beams. The demonstrated results show a reconfigurable photonic router with as many as 16 individually addressable output channels. In addition, multiple channels can be activated at the same time providing an optically induced splitter with configurable outputs as well.

Since the optically induced guiding structure can also be used with wavelengths other than the inducing one~\cite{ref:petter2001}, our technique even allows for the all-optical routing of wavelength-multiplexed signals and the channel addressing can be done using different spectral regions as well. Combined with novel ideas for controlling the trajectory of accelerated beams~\cite{ref:efremidis2011,ref:ye2011,ref:greenfield2011,ref:froehly2011} our presented routing concept can facilitate very complex schemes, and thus we are convinced that this approach for all-optical routing will find many significant applications.

\section*{Acknowledgment}
We acknowledge fruitful discussions about the presented work with Anton~S.~Desyatnikov, Australian National University, Canberra, Australia.

\end{document}